hep-ph/0608053
\documentclass[epj,nopacs]{svjour}
\usepackage{latexsym}
\usepackage{graphicx}
\usepackage{amsmath}
\usepackage{amssymb}
\usepackage{feynmf}
\begin{document}
\title{Quasi-Particle Description of Strongly Interacting Matter: Towards a Foundation}
\author{M. Bluhm\inst{1} \and B. K\"ampfer\inst{1,2} \and R. Schulze\inst{2} \and D. Seipt\inst{2}
}                     
%
%
\institute{Forschungszentrum Rossendorf, PF 510119, 01314 Dresden, Germany \and 
Institut f\"ur Theoretische Physik, TU Dresden, 01062 Dresden, Germany}
\date{Received: date / Revised version: date}
%
\abstract{We confront our quasi-particle model for the equation of state of strongly interacting matter 
with recent first-principle QCD calculations. In particular, we 
test its applicability at finite baryon densities by comparing with Taylor expansion coefficients of the pressure 
for two quark flavours. We outline a chain of approximations starting from the $\Phi$-functional approach to 
QCD which motivates the quasi-particle picture. 
\PACS{
      {PACS-key}{discribing text of that key}   \and
      {PACS-key}{discribing text of that key}
     } 
} 
\maketitle
\section{Introduction}
\label{intro}

In the last years, great progress has been made in the numerical evaluation of QCD thermodynamics from first principles 
(dubbed lattice QCD) even for finite chemical potentials \cite{Allton1,FK,Fodor,deForcrand,Lombardo}. 
While various perturbative expansions \cite{Arnold95,Zhai95,Kajantie03,Vuorinen,Ipp04} 
fail in describing thermodynamics of strongly interacting matter in the vicinity of $T_c$ 
($T_c$ being the (pseudo-) critical temperature of deconfinement and chiral symmetry restauration), 
different phenomenological approaches exist which aim to reproduce the non-perturbative behaviour. For instance, 
models based on quasi-particle pictures with effectively modified properties due to strong interactions are successful 
in describing lattice QCD results \cite{Levai,Schneider,Letessier,Rebhan,Thaler,Ivanov,Peshier}. 
Analytical approaches with a rigorous link to QCD (cf. \cite{Rischke} for a survey) such as direct 
HTL resummation \cite{Andersen,Andersen02b} 
or $\Phi$-functional approach \cite{Blaizot,Blaizot01,Blaizot02} formulated in terms of dressed propagators 
are successful in describing lattice QCD on temperatures $T \gtrsim 2 T_c$. 

It is the aim of the present paper to show the successful applicability of our quasi-particle model (QPM) for 
describing lattice QCD results and to motivate the model starting from the $\Phi$-functional approach to QCD. In 
section \ref{QPM}, we review the QPM and compare with recent lattice QCD results for pressure and entropy density. 
In section \ref{Approxscheme}, a possible chain of approximations is outlined starting from QCD within the 
$\Phi$-functional approximation scheme which motivates our formulation of QCD thermodynamics in terms of quasi-particle 
excitations. We summarize our results in section \ref{cons}. 

\section{QPM and Comparison with Lattice QCD}
\label{QPM}

In our model, the pressure $p$ for $N_f=2$ light quark flavours in thermal equilibrium as function of temperature $T$ 
and one chemical potential $\mu_q$ ($\mu_g=0$) reads 
\begin{equation}
  \label{e:pres0}
  p (T,\mu_q) = \sum_{a = q,g} p_a - B(T,\mu_q) \,,
\end{equation}
where $p_a = d_a/(6 \pi^2) \int_0^\infty dk k^4\left( f_a^+ + f_a^- \right)/\omega_a$ denote the partial pressures 
of quarks ($q$) and transverse gluons ($g$). Here, $d_q=2N_fN_c$, $d_g=N_c^2-1$, $N_c=3$, and 
$f_a^\pm = (\exp( [\omega_a \mp \mu_a]/ T) +S_a)^{-1}$ with $S_{q}=1$ for fermions and $S_g=-1$ for bosons. 
$B(T,\mu_q)$ is determined from thermodynamic self-consistency and the stationarity of $p$ under functional variation 
with respect to the self-energies, $\delta p / \delta \Pi_a = 0$~\cite{Gorenstein}. $\Pi_a$ enter the quasi-particle 
dispersion relations $\omega_a$ being approximated by asymptotic mass shell expressions near the light cone, 
$\omega_a = \sqrt{k^2 + \Pi_a}$. We employ the asymptotic expressions of the gauge independent hard thermal 
(dense) loop self-energies \cite{LeBellac96}. Finite bare quark masses $m_{0;q}$ as used in lattice 
simulations can be implemented following \cite{Pisarski}. 

By replacing the running coupling $g^2$ in $\Pi_a$ with an effective coupling $G^2(T,\mu_q)$, non-perturbative 
effects in the vicinity of $T_c$ are accomodated. In this way, we achieve enough flexibility to describe lattice 
QCD results. We \linebreak parametrize $G^2(T,\mu_q=0)$ (cf. \cite{Blu05} for details) by 
\begin{equation}
  \label{e:G2param}
  \hspace{-0.5mm}G^2(T,\mu_q=0) = \left\{
    \begin{array}{l}
      G^2_{\rm (2)} (\xi(T)), \,T \ge T_c,
      \\[3mm]
      G^2_{\rm (2)}(\xi(T_c)) + b (1- \frac{T}{T_c}), \, T < T_c ,
    \end{array}
  \right.
\end{equation}
where $G^2_{\rm (2)}$ is the relevant part of the 2-loop running coupling and 
$\xi(T) = \lambda (T - T_s)/T_c$ contains a scale parameter $\lambda$ and an infrared regulator 
$T_s$. The effective coupling $G^2$ for arbitrary $T$ and $\mu_q$ can be found by solving a 
quasi-linear partial differential equation which follows from Maxwell's relation, 
\begin{equation}
  \label{equ:flow}
  a_{\mu_q}\frac{\partial G^2}{\partial\mu_q} + a_T\frac{\partial G^2}{\partial T} = b \,.
\end{equation}
The coefficients in (\ref{equ:flow}) explicitly read (neglecting for simplicity additional contributions stemming 
from $T$-\linebreak 
dependent bare quark masses as employed in lattice simulations) 
\begin{eqnarray}
  a_T & = & {\rm I_1} \frac{C_f}{4}\left(T^2+\frac{\mu_q^2}{\pi^2}\right) \,,\\
  a_{\mu_q} & = & -{\rm I_2} \frac{C_f}{4}\left(T^2+\frac{\mu_q^2}{\pi^2}\right) \\ \nonumber & & - 
  {\rm I_3}\left(\left[N_c+\frac{N_f}{2}\right]\frac{T^2}{6} + \frac{N_cN_f}{12\pi^2}\mu_q^2\right) \,,\\
  b & = & - {\rm I_1} \frac{C_f}{2}TG^2 + {\rm I_2} \frac{C_f}{2}\frac{\mu_q}{\pi^2}G^2 \\ 
  \nonumber & & + 
  {\rm I_3} \frac{N_cN_f}{6\pi^2}\mu_q G^2 \,.
\end{eqnarray}
Here, 
\begin{eqnarray}
  {\rm I_1} & = & \frac{d_q}{4\pi^2T}\int_0^\infty d k \frac{k^2}{\omega_q}\bigg(e^{\omega_q^+/T}(f_q^-)^2 
  - e^{\omega_q^-/T}(f_q^+)^2\bigg) , \\ \nonumber 
  {\rm I_2} & = & \frac{d_q}{2\pi^2T}\int_0^\infty d k \frac{k^2}{\omega_q} \bigg( f_q^+ - \frac{L_q f_q^+}{\omega_q} 
  \bigg[\frac{1}{\omega_q} + \frac{e^{\omega_q^-/T}}{T}f_q^+\bigg] \\ 
  & & \hspace{-4mm} + \frac{\mu_q}{2T}e^{\omega_q^-/T}(f_q^+)^2 + f_q^- - \frac{L_q f_q^-}{\omega_q} \bigg[\frac{1}{\omega_q} + 
  \frac{e^{\omega_q^+/T}}{T}f_q^-\bigg] \\ \nonumber & & \hspace{-4mm} - \frac{\mu_q}{2T}e^{\omega_q^+/T}(f_q^-)^2 \bigg) , \\
  {\rm I_3} & = & \frac{d_g}{\pi^2T}\int_0^\infty d k \frac{k^2}{\omega_g}f_g\left(1-\frac{L_g}{\omega_g}\left[\frac{1}{\omega_g} + 
      \frac{e^{\omega_g/T}}{T}f_g\right]\right) 
\end{eqnarray}
with $f_g^\pm\equiv f_g$, $\omega_q^\pm=\omega_q\pm\mu_q$, $L_a=2k^2/3+\Pi_a/2$ and $C_f=(N_c^2-1)/2N_c$. 

\begin{figure}[hbt]
  \centering{
  \includegraphics[scale=0.23,angle=-90.]{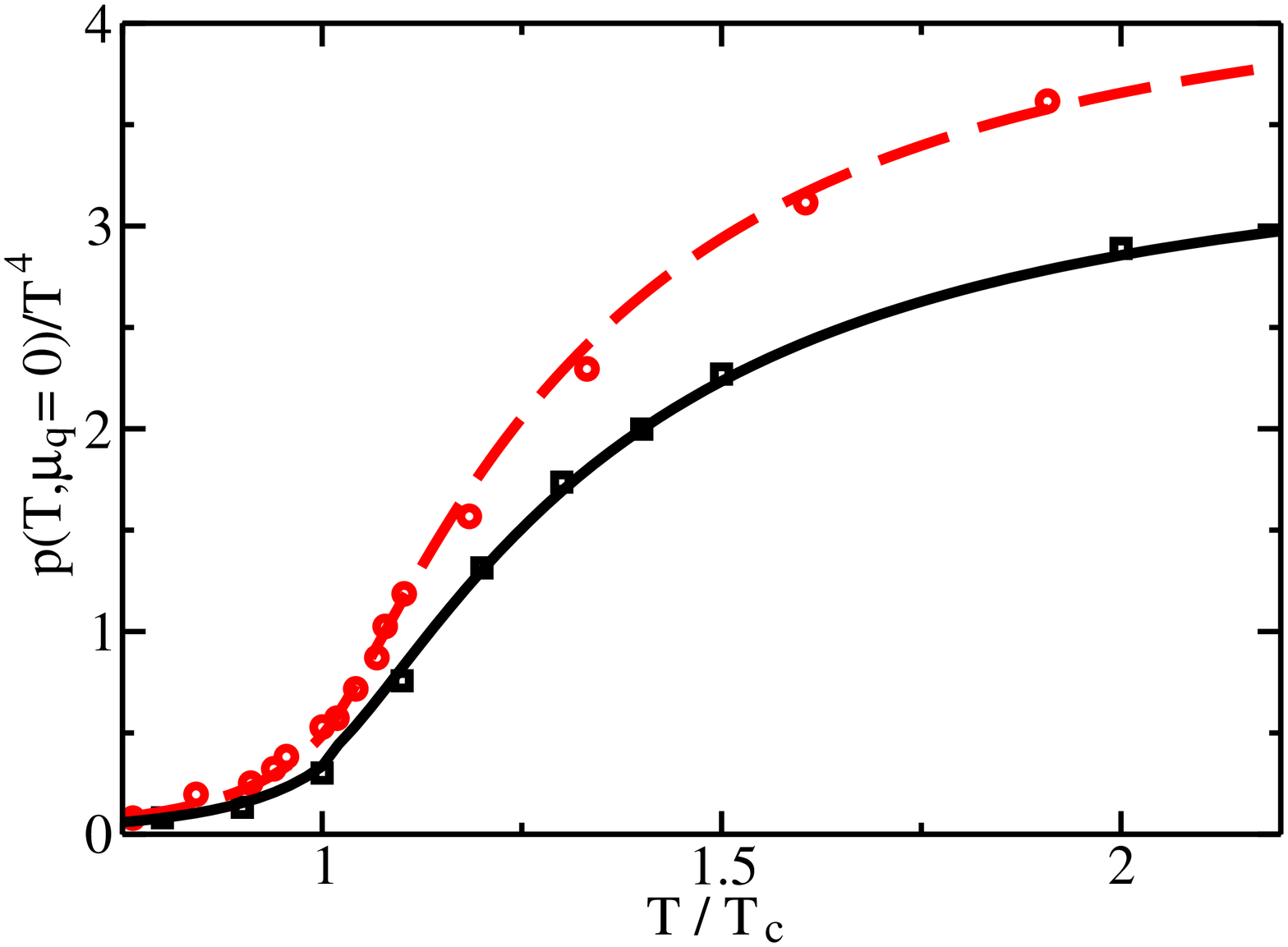}\\
  \includegraphics[scale=0.23,angle=-90.]{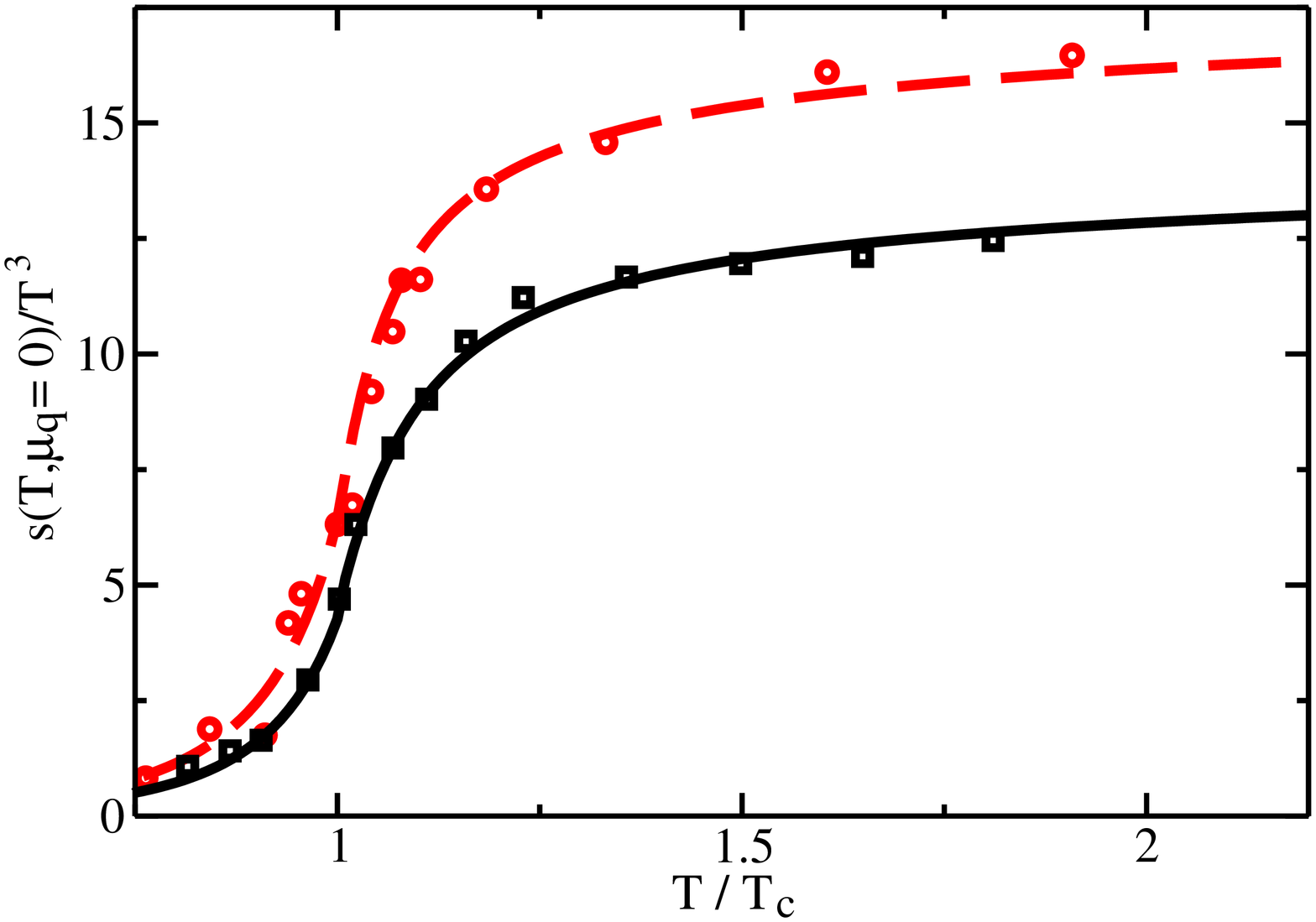}}
  \caption{Comparison of our QPM with lattice QCD results (symbols) for $p(T,\mu_q=0)/T^4$ (upper panel) and 
    $s(T,\mu_q=0)/T^3$ (lower panel) as functions of $T/T_c$ for $N_f=2$ (squares) \cite{Kar1} and $N_f=2+1$ 
    ($\mu_s=0$) (circles) \cite{Kar1,Kar2}. Raw lattice QCD data are continuum extrapolated as advocated in \cite{Kar1,Kar3}. 
    QPM parameters: $\lambda = 4.4$, $T_s=0.67T_c$, $b=344.4$, 
    $B(T_c)=0.31 T_c^4$ with $T_c=175$ MeV for $N_f=2$ and $\lambda = 7.6$, $T_s=0.80T_c$, $b=348.7$, 
    $B(T_c)=0.52 T_c^4$ with $T_c=170$ MeV for $N_f=2+1$. \label{fig:1}}
\end{figure}

Entropy density $s=\partial p/\partial T=\sum_{a=q,g} s_a$ and net density $n=n_q=\partial p/\partial\mu_q$ 
follow from (\ref{e:pres0}) as 
\begin{eqnarray}
  \label{equ:entrcomp}
  s_a & = & \frac{d_a}{2\pi^2 T} \int_0^\infty d k k^2 \bigg(\frac{\frac{4}{3}k^2 + \Pi_a}{\omega_a}\left[f_a^+ + f_a^-\right] \\ 
  \nonumber 
    & & - \mu_a\left[f_a^+ - f_a^-\right]\bigg) \,, \\ 
    \label{equ:denscomp}
    n_q & = & \frac{d_q}{2\pi^2}\int_0^\infty d k k^2 \left[f_q^+-f_q^-\right] \,.
\end{eqnarray}
In Fig. \ref{fig:1}, we exhibit QPM results for $p$ and $s$ at $\mu_q=0$ compared with lattice QCD results for different numbers of quark 
flavours \cite{Kar1,Kar2}. 

\begin{figure}[hbt]
  \centering{
  \includegraphics[scale=0.23,angle=-90.]{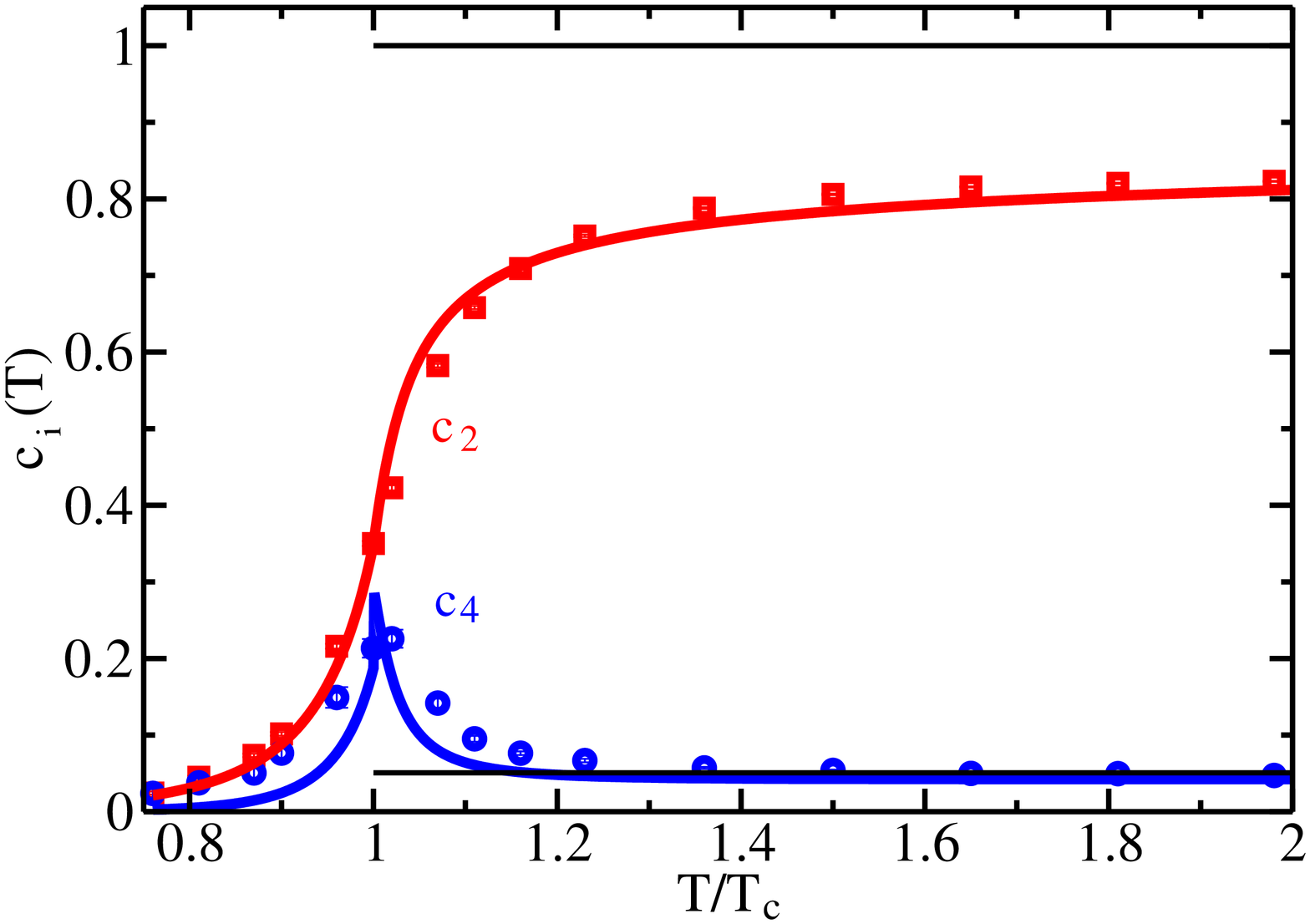}\\
  \includegraphics[scale=0.23,angle=-90.]{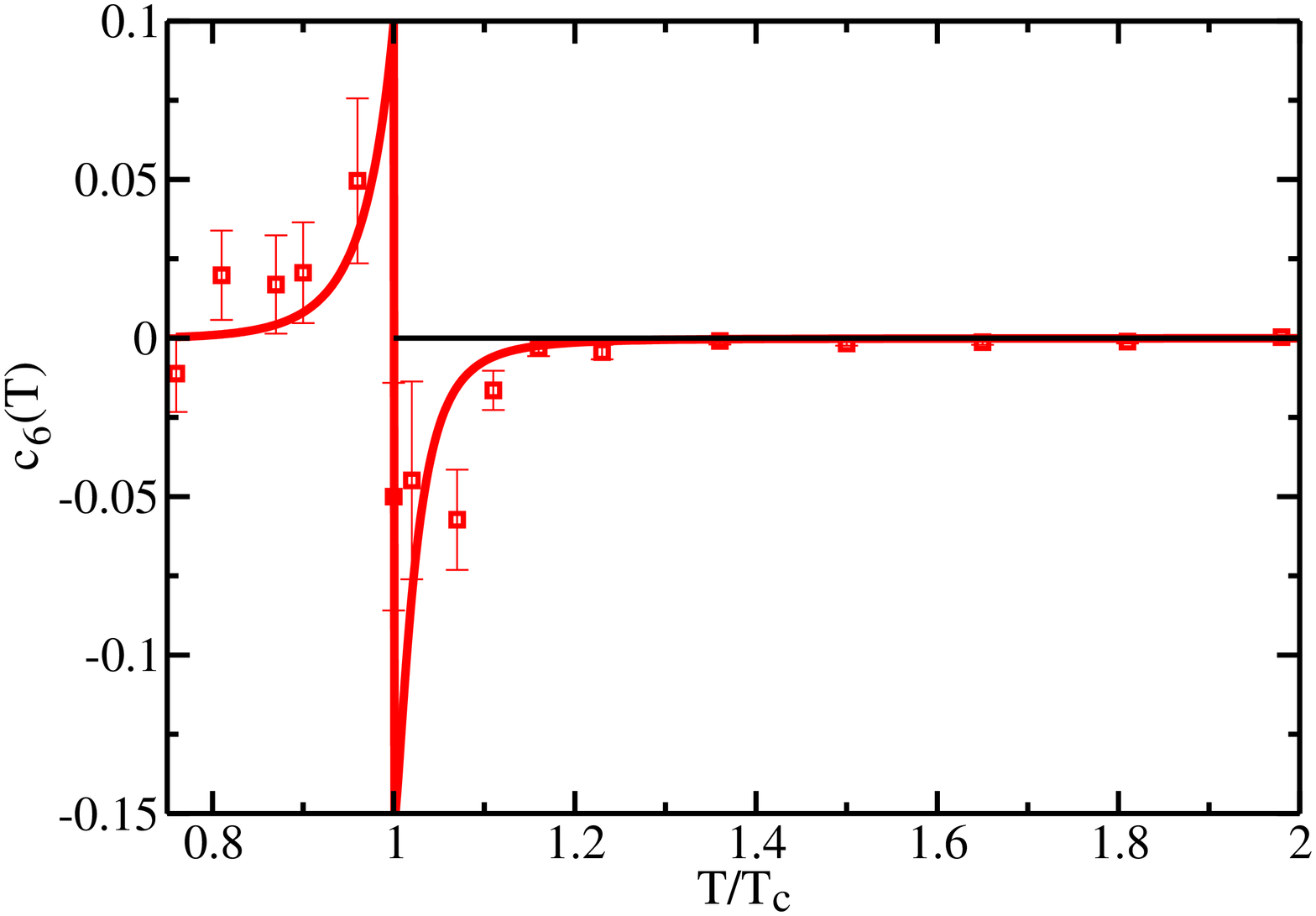}}
  \caption{Comparison of our QPM with lattice QCD results (symbols) \cite{All05} for $c_{2,4}(T)$ (upper panel) 
    and $c_6(T)$ (lower panel) as functions of $T/T_c$ for $N_f=2$. 
    QPM parameters: $\lambda = 12.0$, $T_s=0.87T_c$, $b=426.1$, with $T_c=175$ MeV. The horizontal lines at $T\ge T_c$ depict 
    the corresponding Stefan-Boltzmann values highlighting the effects of strong interaction near $T_c$. \label{fig:2}}
\end{figure}

Recently, the decomposition of $p$ into a Taylor series in powers of ($\mu_q/T$) for small $\mu_q$ was studied in 
lattice QCD \cite{All05}, 
\begin{equation}
  \label{e:presdecomp}
  p(T, \mu_q) = T^4 \sum_{n=0}^\infty c_n(T) \left( \frac{\mu_q}{T} \right)^n \,.
\end{equation}
The expansion coefficients $c_n(T)$, vanishing for odd $n$ and depending only on temperature $T$, follow using (\ref{e:pres0}) from 
\begin{equation}
  \label{e:coeff1}
  c_n(T) = \left.\frac{1}{n!} \frac{\partial^n (p/T^4)}{\partial  (\mu_q/T)^n}\right|_{\mu_q = 0} \,.
\end{equation}
$c_n(T)$ depend on $G^2$ and its derivatives with respect to $\mu_q$ at $\mu_q=0$, thus testing (\ref{equ:flow}). 
Furthermore, net density $n$ can 
also be decomposed into a Taylor series at small $\mu_q$ with expansion coefficients $c_n(T)$. Therefore, the higher order 
coefficients $c_{2,4,6}(T)$ serve for a more direct test of the applicability of our model at finite $\mu_q$. In Fig. 
\ref{fig:2}, we compare $c_{2,4,6}(T)$ evaluated from (\ref{e:coeff1}) with lattice QCD results for $N_f=2$ \cite{All05}. 
In particular, the pronounced structures in the vicinity of $T_c$ are fairly well reproduced (cf. \cite{Blu05}). 

\section{Foundations of the QPM}
\label{Approxscheme}

Having successfully reproduced first-principle lattice QCD results, 
it would be desirable to establish contact between our {\it ad hoc} introduced QPM in section~\ref{QPM} and QCD 
as the fundamental microscopic gauge field theory of strong interactions. In order to motivate our quasi-particle
model, we present a possible chain of approximations starting from QCD within the $\Phi$-functional 
approach following the pioneering work \cite{Blaizot,Blaizot01,Blaizot02}. 
We concentrate on entropy density $s$ and net density $n$, as they turn out to possess a simple structure 
supporting the picture of quasi-particle excitations. Other thermodynamic quantities such as pressure $p$ or energy
density $e$ are determined from $s$ and $n$. Although rather strong 
assumptions become mandatory in the derivation, one should be aware of the remarkable success of our QPM 
in describing lattice QCD results. 

In the $\Phi$-functional approach \cite{Luttinger60,Baym62} to QCD, the thermodynamic potential 
$\Omega=-T\ln Z$ can be expressed as a functional of dressed propagators of gluons $D$, quarks $S$ and 
Faddeev-Popov ghost fields $G$, 
\begin{eqnarray}
  \label{equ:phi1}
  \hspace{-4mm}\frac{\Omega [D,S,G]}{T} = & \frac{1}{2}{\rm Tr}[\ln D^{-1}\!-\!\Pi D] -
  {\rm Tr}[\ln S^{-1}\!-\!\Sigma S] \\ \nonumber & \hspace{-4mm}-{\rm Tr}[\ln
  G^{-1}-\Xi G] + \Phi[D,S,G]\,. 
\end{eqnarray}
Here, ghost field contributions compensate for possible unphysical degrees of freedom in the gluon propagator. 
While the propagators in~(\ref{equ:phi1}) depend on the specific gauge, $\Omega=-pV$ must be gauge independent. 
For convenience, we choose the Coulomb gauge in the following in which ghost fields do not propagate 
and the gluon propagator consists only of the physical transverse and longitudinal modes. The functional 
$\Phi[D,S]$ is given by the infinite sum of all $2\,$-particle irreducible skeleton diagrams constructed 
from $D$ and $S$. 

The self-energies are related to the dressed propagators by Dyson's equations 
\begin{equation}
  \label{equ:phi2}
  \Pi [D] = D^{-1} - D_0^{-1}\,\,,\,\,\Sigma [S] = S^{-1} - S_0^{-1}\,, 
\end{equation}
where $D_0$ and $S_0$ represent the bare propagators of gluon and
quark fields, respectively. Demanding the stationarity of $\Omega$ under functional 
variation with respect to the dressed propagators~\cite{Lee60} 
\begin{equation}
  \label{equ:phi4}
  \left.\frac{\delta\Omega [D,S]}{\delta D}\right|_{D_0} = \left.\frac{\delta\Omega
  [D,S]}{\delta S}\right|_{S_0} = 0\,,
\end{equation}
the self-energies follow self-consistently by cutting a
dressed propagator line in $\Phi$ resulting in the gap equations 
\begin{equation}
  \label{equ:phi5}
  \Pi = 2\frac{\delta\Phi [D,S]}{\delta D}\,\,,\,\,\Sigma = -\frac{\delta\Phi [D,S]}{\delta S}\,.
\end{equation}

The trace $"{\rm Tr}"$ in~(\ref{equ:phi1}) has to be taken over all states of the relativistic many-particle system. 
In the imaginary time formalism it can be rewritten in the form ${\rm Tr} \rightarrow {\rm tr} 
\,\beta VT\sum_{n=-\infty}^{+\infty}\int d^3k/(2\pi)^3$. Here, $V$ is the volume of the system, $\beta=1/T$ 
and $"{\rm tr}"$ denotes the remaining trace over occuring discrete indices including colour, 
flavour, Lorentz or spinor indices. Introducing the four-momentum
$k^\nu=(\omega,\vec{k})=(i\omega_n,\vec{k})$, the sums have to be taken
over the Matsubara frequencies $\omega_n = 2n\pi T$ (or $(2n+1)\pi T -i\mu$) for gluons (or quarks). 
They can be evaluated by using standard contour integration techniques in the complex $\omega$-plane 
\cite{LeBellac96,Kapusta89} wrapping up the poles of the propagators. Expressing the analytic propagators in terms 
of their spectral densities $\rho$, one can define 
\begin{equation}
  \rho_{D(S)}(\omega, |\vec{k}|) = 2 \lim_{\epsilon\rightarrow 0} {\rm Im}\,D(S) (\omega +i\epsilon, |\vec{k}|) 
\end{equation}
for real $\omega$. Similarly, the imaginary parts of functions of the analytic propagators obeying the same pole structures 
can be defined. Hence, $\Omega$ reads with retarded propagators $D$ and $S$ depending on $\omega$ and 
$k=|\vec{k}|$ 
\begin{eqnarray}
  \label{equ:phi7} 
  \frac{\Omega [D,S]}{V} = & {\rm
  tr} \int \frac{d^4k}{(2\pi)^4} n(\omega)\, {\rm Im} [\ln D^{-1}-\Pi\, D] \\ \nonumber 
  & \hspace{-1.8cm}+ 2{\rm tr} \int \frac{d^4k}{(2\pi)^4} f(\omega)\, {\rm Im} [\ln
  S^{-1}-\Sigma\, S] + \frac{T}{V}\Phi[D,S]\,, 
\end{eqnarray}
where $\int d^4k=\int d^3k\int d\omega$, and 
$n(\omega)=(e^{\beta\omega}-1)^{-1}$ ($f(\omega)=(e^{\beta(\omega-\mu)}+1)^{-1}$) denotes the statistical distribution 
function for gluons (quarks with chemical potential $\mu$). 

Due to the stationarity property (\ref{equ:phi4}), entropy density $s=-\partial (\Omega/V)/\partial T$ and 
net density $n=-\partial (\Omega/V)/\partial\mu$ contain 
only explicit temperature and chemical potential 
derivatives of $n(\omega)$ and $f(\omega)$, although the propagators in (\ref{equ:phi7}) 
depend implicitly on $T$ and $\mu$ through their spectral densities. 
Using ${\rm Im}\,(\Pi D) = {\rm Im}\,\Pi\, {\rm Re}\,D + {\rm Re}\,\Pi\,{\rm Im}\,D$, one finds for the entropy density 
$s = s_g + s_q + s'$ with 
\begin{equation}
  \label{equ:phi10}
  s_g = -{\rm tr}\int\frac{d^4k}{(2\pi)^4}\frac{\partial
  n(\omega)}{\partial T} [{\rm Im}\ln D^{-1}-{\rm Im}\,\Pi\,{\rm Re}
  D]\,, 
\end{equation}
\begin{equation}
  \label{equ:phi11}
  s_q = -2{\rm tr}\int\frac{d^4k}{(2\pi)^4}\frac{\partial
  f(\omega)}{\partial T} [{\rm Im}\ln S^{-1}-{\rm Im}\,\Sigma\,{\rm
  Re} S]\,,
\end{equation}
\begin{eqnarray}
  \label{equ:phi12}
  \nonumber 
  \hspace{-5mm}
  s' & = - \left.\frac{\partial(\frac{T}{V}\Phi[D,S])}{\partial T}\right|_{D,S} + 2{\rm tr}\int\frac{d^4k}{(2\pi)^4}\frac{\partial
  f(\omega)}{\partial T} {\rm Re}\Sigma\,{\rm Im} S \\ 
   & + {\rm tr}\int\frac{d^4k}{(2\pi)^4}\frac{\partial n(\omega)}{\partial T} {\rm Re}\Pi\,{\rm Im} D \,. 
\end{eqnarray}
Similarly, for the net density one finds $n = n_q + n'$ with 
\begin{eqnarray}
  \label{equ:phi11+}
  n_q & = & -2{\rm tr}\int\frac{d^4k}{(2\pi)^4}\frac{\partial
  f(\omega)}{\partial \mu} [{\rm Im}\ln S^{-1}-{\rm Im}\,\Sigma\,{\rm
  Re} S] , \\ 
  \label{equ:phi12+}
  n' & = & 2{\rm tr}\int\frac{d^4k}{(2\pi)^4}\frac{\partial
  f(\omega)}{\partial \mu} {\rm Re}\Sigma\,{\rm Im} S \\ \nonumber
  & & - \left.\frac{\partial(\frac{T}{V}\Phi[D,S])}{\partial \mu}\right|_{D,S} \,. 
\end{eqnarray}
While the sum integrals in $\Omega$ (\ref{equ:phi7}) contain ultraviolet divergencies which must be regularized, the 
expressions for $s_g$, $s_q$ and $n_q$ in~(\ref{equ:phi10},~\ref{equ:phi11}) and~(\ref{equ:phi11+}) 
are manifestly ultraviolet convergent because the derivatives of the statistical distribution 
functions vanish for $\omega\rightarrow\pm\infty$. In addition, introducing real multiplicative 
renormalization factors for propagators and self-energies, these factors simply drop out of $s$ and $n$. 

Self-consistent (or $\Phi$-derivable) approximation schemes preserve the stationarity property 
(\ref{equ:phi4}) of $\Omega$ when 
truncating the infinite sum in $\Phi$ at a specific loop order while corresponding self-energies and propagators 
are self-\linebreak
consistently evaluated from (\ref{equ:phi5}) and Dyson's equations. Nevertheless, self-consistency 
does not guarantee gauge invariance which is an important issue in truncated expansion schemes. In fact, by modifying 
propagators but leaving vertices unaffected Ward identities are violated. 

We consider $\Phi$ at 2 - loop order in the following which is diagrammatically represented by \cite{Peshier01} 
\vspace{.3cm}
\begin{fmffile}{phis}
\begin{equation}
  \vspace{.5cm}
  \Phi = \frac{1}{12} \,\,\,\parbox{25mm}{\begin{fmfgraph}(39,25) \fmfleft{i}\fmfright{o} \fmf{phantom,tension=1.9}{i,v1}
      \fmf{phantom,tension=0.13}{v1,v2} \fmf{phantom,tension=0.4}{v2,o} \fmffreeze 
      \fmf{photon,right,tension=4}{v1,v2,v1} \fmf{photon,tension=1}{v1,v2} 
      \fmfdot{v1,v2} \end{fmfgraph}} \hspace{-1.3cm} + \,\frac{1}{8} \,\,\,
  \parbox{25mm}{\begin{fmfgraph}(55,25) \fmfleft{i}\fmfright{o} \fmf{phantom,tension=1.8}{i,v1}
      \fmf{phantom,tension=0.08}{v1,v2} \fmf{phantom,tension=0.08}{v2,v3} 
      \fmf{phantom,tension=0.8}{v3,o} \fmffreeze \fmf{photon,right,tension=1}{v2,v1,v2} \fmf{photon,right,tension=1}{v2,v3,v2} 
      \fmfdot{v2} \end{fmfgraph}} 
  \hspace{-5mm} - \,\,\frac{1}{2} \hspace{-1mm}
  \parbox{20mm}{\begin{fmfgraph}(55,25) \fmfleft{i}\fmfright{o} \fmf{phantom,tension=0.4}{i,v1}
      \fmf{phantom,tension=0.13}{v1,v2} \fmf{phantom,tension=0.4}{v2,o} \fmffreeze 
      \fmf{vanilla,right,tension=0.6}{v1,v2,v1} \fmf{photon,tension=1}{v1,v2} 
      \fmfdot{v1,v2} \end{fmfgraph}} \hspace{-3mm}.
\end{equation}
\end{fmffile}
\hspace{-1mm}Here, wiggly (solid) lines denote gluons (quarks). The self-consistent self-energies are accordingly 
\begin{fmffile}{selfs}
\vspace{.3cm}
\begin{equation}
  \vspace{.5cm}
  \Pi = \frac{1}{2} \,\,\parbox{25mm}{\begin{fmfgraph}(50,25) \fmfleft{i} \fmfright{o} \fmf{photon,tension=1}{i,v1} 
      \fmf{photon,right,tension=0.3}{v1,v2} 
      \fmf{photon,right,tension=0.3}{v2,v1}\fmf{photon}{v2,o} \fmfdot{v1,v2} \end{fmfgraph}} \hspace{-6.5mm}
  + \frac{1}{2} \,\,\parbox{25mm}{\begin{fmfgraph}(40,25) \fmfleft{i} \fmfright{o} \fmf{photon,tension=1}{i,v1} \fmf{photon}{v2,o}
      \fmf{phantom,tension=50}{v1,v2} \fmf{photon,left=120}{v1,v2}\fmfdot{v1} \end{fmfgraph}} \hspace{-1cm} - \,
  \parbox{25mm}{\begin{fmfgraph}(52,25) \fmfleft{i} \fmfright{o} \fmf{photon,tension=1}{i,v1} 
      \fmf{vanilla,right,tension=0.3}{v1,v2,v1} \fmf{photon}{v2,o} \fmfdot{v1,v2} \end{fmfgraph}} \hspace{-5mm},
\end{equation}
\end{fmffile}
\vspace{.1cm}
\begin{fmffile}{newself}
\begin{equation}
\vspace{.3cm}
  \Sigma = \,\,\parbox{25mm}{\begin{fmfgraph}(50,25) \fmfleft{i} \fmfright{o} \fmf{vanilla,tension=3.}{i,v1} 
      \fmf{vanilla,tension=0.12}{v1,v2} 
      \fmf{vanilla,tension=3.}{v2,o} \fmf{photon,left,tension=1}{v1,v2} 
      \fmfdot{v1,v2} \end{fmfgraph}} \hspace{-4mm}.
\end{equation}
\end{fmffile}
\hspace{-2mm}Although vertex corrections can be implemented self-\linebreak
consistently \cite{Freedman78}, they turn out to be negligible 
at 2 - loop order in $\Phi$ \cite{Blaizot01}. In addition, $s'=n'=0$ is found for the residual contributions of 
entropy density and net density in (\ref{equ:phi12}) and (\ref{equ:phi12+}) at 2 - loop order 
\cite{Blaizot01}. This topological feature being 
related to (\ref{equ:phi5}) has also been observed in massless $\Phi^4$-theory 
\cite{Peshier01,Peshier98a} and in QED \cite{Vanderheyden}. 

Concentrating on the gluonic contribution $s_g$,~(\ref{equ:phi10}) can be rewritten by using the identity 
\begin{eqnarray}
  \label{equ:phi18}
  {\rm Im}[\ln D^{-1}(\omega,k)] = - \pi{\rm sgn}(\omega)\,\Theta(-{\rm Re} D^{-1}(\omega,k)) \\ \nonumber 
  + \arctan\left(\frac{{\rm Im} \Pi(\omega,k)}{{\rm Re} D^{-1}(\omega,k)}\right) 
\end{eqnarray}
where $-\pi/2<\arctan x<\pi/2$. Hence, $s_g$ can be decomposed into $s_g = s_{g,QP} + s_{g,LD}$ with 
\begin{eqnarray}
  \label{equ:phi20}
  \hspace{-6mm}s_{g,QP} & = & {\rm tr} 
  \int\frac{d^3k}{(2\pi)^3}\int\frac{d\omega}{2}\frac{\partial
  n(\omega)}{\partial T}{\rm sgn}(\omega)\,\Theta(-{\rm
  Re} D^{-1})\, , \\ 
  \label{equ:phi21}
  \hspace{-6mm}s_{g,LD} & = & {\rm tr} \int\frac{d^4k}{(2\pi)^4}\frac{\partial
  n(\omega)}{\partial T}\bigg\{{\rm Im} \Pi\,{\rm Re} D \\ \nonumber 
  & & - \arctan\left(\frac{{\rm Im} \Pi}{{\rm Re} D^{-1}}\right)\bigg\} \, .
\end{eqnarray}
Here, (\ref{equ:phi20}) accounts for the contribution of dynamical \linebreak quasi-particles to $s_g$ defined by the poles 
of $D$ and (\ref{equ:phi21}) represents the contribution from the continuum part of the spectral density associated 
with a cut below the light cone $|\omega|<k$ \cite{Pisarski,Peshier04} representing Landau damping. 
Applying a similar identity for ${\rm Im}[\ln S^{-1}(\omega,k)]$, $s_q$ and $n_q$ 
in (\ref{equ:phi11},~\ref{equ:phi11+}) can be decomposed similarly into quasi-particle and Landau-damping contributions. 

\begin{figure}[hbt]
  \centering{
  \includegraphics[scale=0.25,angle=-90.]{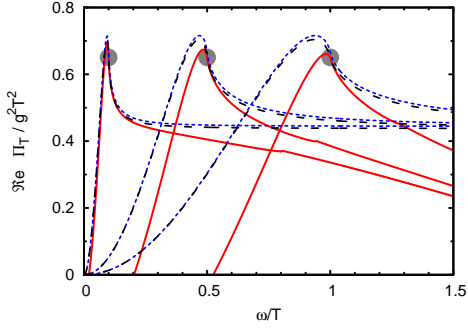}}
  \caption{Comparison of HTL approximation (the nearly indistinguishable short and long dashed curves are for massless (\ref{equ:phi13}) 
    (cf. \cite{Thoma}) and massive quarks, $m_q=0.4\,T$, as used in lattice calculations \cite{Kar1,Kar2,Kar3,All05}, 
    respectively) with 1-loop results 
    (solid curves for massive quarks) \cite{Kalashnikov} of the scaled real part of the 
    transverse gluon self-energy as function of $\omega /T$ for $k/T=0.1$, $0.5$ and $1$ from left to right. $N_f=2$, 
    $\mu_i=0$. In the vicinity of the light cone $\omega = k$ (shaded regions), 
    HTL results approximate 1-loop results fairly well even for $k\sim T$ which is the thermodynamically relevant part 
    in momentum space but with decreasing 
    agreement for increasing $k$. \label{fig:3}}
\end{figure}

In Coulomb gauge, $D$ consists of a longitudinal and a transverse part, $D_L$ and $D_T$. 
Similarly, the (massless) quark propagator consists of two different branches with chirality either equal (positive 
energy states) or opposite (negative energy states) to helicity. 
By employing the gauge invariant hard thermal loop (HTL) expressions 
$\hat{\Pi}$ ($\hat{\Sigma}$) for the gluon (quark) self-energies in the following, one obtains gauge invariant 
approximations of $s$ and $n$. The HTL expressions read \cite{LeBellac96} 
\begin{eqnarray}
  \label{equ:phi13-}
  \hat{\Pi}_L(\omega,k) & = & \hat{m}_D^2\left(1-\frac{\omega}{2k} \ln \frac{\omega+k}{\omega-k}\right) ,\\
  \label{equ:phi13}
  \hat{\Pi}_T(\omega,k) & = & \frac{1}{2}\left(\hat{m}_D^2 +
  \frac{\omega^2-k^2}{k^2}\hat{\Pi}_L(\omega,k)\right) ,\\
  \label{equ:phi14}
  \hat{\Sigma}_\pm(\omega,k) & = & \frac{\hat{M}^2}{k}\left(1 - 
  \frac{\omega \mp k}{2k}\ln\frac{\omega+k}{\omega-k}\right) ,
\end{eqnarray}
with Debye screening mass (allowing, in general, for different chemical potentials $\mu_i$) 
\begin{equation}
  \label{equ:phi14+}
  \hat{m}_D^2 = \left([2N_c+N_f]T^2 + N_c\sum_i \frac{\mu_i^2}{\pi^2}\right)\frac{g^2}{6} \,,
\end{equation}
long-wavelength fermionic frequency 
\begin{equation}
  \label{equ:phi14++}
  \hat{M}^2 = \frac{N_c^2-1}{16 N_c}\left(T^2 + \frac{\mu_i^2}{\pi^2}\right)g^2 \,,
\end{equation}
and running coupling $g^2$. Although being derived originally for soft external momenta $\omega, k\sim gT \ll T$, 
they coincide on the light cone with complete 1-loop results \cite{Thoma} as exhibited in Fig.~\ref{fig:3}. Finite 
quark masses, $m<T$, turn out to be negligible. The corresponding propagators are evaluated from Dyson's equations. 

For $k\sim T, \mu$ the poles of both, longitudinal gluon propagator as well as 
abnormal fermion branch have exponentially vanishing residues \cite{Pisarski} giving only minor contributions 
to the thermodynamics. Therefore, we assume that these collective modes can be neglected in the following. Furthermore, 
being a severe approximation, we also neglect any imaginary parts of the self-energies, i.~e. 
${\rm Im} \hat{\Pi}_T = {\rm Im} \hat{\Sigma}_+ = 0$. Then, the Landau damping contributions to $s_g$, $s_q$ and $n_q$ 
vanish. Finite width effects associated with imaginary parts of the self-energies are discussed by 
Peshier \cite{Peshier04}. Including Landau damping as well as the exponentially suppressed modes, it 
was shown in \cite{Rebhan} that in this way some ambiguities arising when solving (\ref{equ:flow}) can 
be eliminated. 

Performing the $\omega$-integration in~(\ref{equ:phi20}) (but now for $\hat{D}_T$), the only contributions stem from 
$\omega^2 \ge \omega_T^2$ because of the $\Theta$-function, where $\omega_T$ is the positive solution of 
$\omega^2 - k^2 - \hat{\Pi}_T(\omega,k) = 0$. Therefore, the $\omega$-integral in~(\ref{equ:phi20}) reads 
\begin{eqnarray}
  \label{equ:phi22}
  \int_{-\infty}^\infty\frac{d\omega}{2}\frac{\partial
  n(\omega)}{\partial T}{\rm sgn}(\omega)\,
  \Theta(-{\rm Re}\hat{D}_T^{-1}) = \\ \nonumber 
  \int_\infty^{\omega_{T}}\frac{d
  \omega}{2}\left(\frac{\partial n(-\omega)}{\partial T} - \frac{\partial n(\omega)}{\partial T}\right) \,.
\end{eqnarray}
The remaining integration is performed through an integration by parts using 
$-\partial n(\omega)/\partial T = \partial n(-\omega)/\partial T = \partial \sigma(\omega)/\partial\omega$ for 
the spectral function $\sigma(\omega)=-n(\omega)\ln n(\omega)+\left(1+n(\omega)\right)\ln\left(1+n(\omega)\right)$. 
Taking the trace over polarization and colour degrees of freedom for the transverse gluon modes, one finds 
\begin{eqnarray}
  \label{equ:phi27}
  s_{g,QP}  =  
  - 2(N_c^2-1)\int\frac{d^3k}{(2\pi)^3}\bigg(\ln(1-e^{-\beta\omega_{T}}) \\ 
    \nonumber - \frac{\beta\omega_{T}}{e^{\beta\omega_{T}}-1}\bigg) \,.
\end{eqnarray}

Similarly, $s_{q,QP}$ can be evaluated, where non-vanishing contributions to the $\omega$-integration stem from 
$\omega \ge \omega_+$. Here, $\omega_+$ is the solution of $\omega - k - \hat{\Sigma}_+(\omega,k) = 0$ for 
the positive fermion branch. Using $-\partial f(\omega)/\partial T = \partial \sigma(\omega)/\partial\omega$ for the 
spectral function $\sigma(\omega)=-f(\omega)\ln f(\omega)-(1-f(\omega))\ln(1-f(\omega))$, the $\omega$-integral 
can be integrated by parts. Antiquarks are included by simply replacing $\mu \rightarrow -\mu$ in $f(\omega)$. 
Taking the trace over remaining spin, colour and flavour degrees of freedom, one finds 
\begin{eqnarray}
  \label{equ:phi28} 
  s_{q,QP} & = 
  2N_cN_f\int\frac{d^3k}{(2\pi)^3}\bigg(\ln(1+e^{-\beta(\omega_{+}-\mu)}) \\ \nonumber & +
  \frac{\beta(\omega_{+}-\mu)}{e^{\beta(\omega_{+}-\mu)}+1}\bigg)
  \\ \nonumber & 
  \quad\!+
  2N_cN_f\int\frac{d^3k}{(2\pi)^3}\bigg(\ln(1+e^{-\beta(\omega_{+}+\mu)})
  \\ \nonumber & +
  \frac{\beta(\omega_{+}+\mu)}{e^{\beta(\omega_{+}+\mu)}+
  1}\bigg) \,. 
\end{eqnarray}
$s_{g,QP}$ and $s_{q,QP}$ in (\ref{equ:phi27},~\ref{equ:phi28}) represent entropy density contributions of non-interacting 
quasi-particles with quantum numbers of transverse gluons (quarks) and dispersion relation $\omega_T$ ($\omega_+$). 

Correspondingly, $n_{q,QP}$ is evaluated using \linebreak 
$-\partial f(\omega)/\partial \mu = \partial f(\omega)/\partial\omega$. 
Adding antiquarks by $\mu \rightarrow -\mu$ in $f(\omega)$ (note, now 
$\partial f(\omega)/\partial \mu = \partial f(\omega)/\partial\omega$) and taking the trace, $n_{q,QP}$ reads 
\begin{eqnarray}
  \label{equ:phi28+} 
  \hspace{-7mm}n_{q,QP} & = 
  2N_cN_f\int\frac{d^3k}{(2\pi)^3}\bigg(\frac{1}{e^{\beta(\omega_{+}-\mu)}+1} - \frac{1}{e^{\beta(\omega_{+}+\mu)}+1}\bigg) \,. 
\end{eqnarray}
Finally, we approximate the quasi-particle dispersion relations by the asymptotic mass shell expressions near the light 
cone, thus neglecting any momentum or energy dependence of the self-energies. We employ 
$\omega_{T}\rightarrow\omega_g=\sqrt{k^2+\Pi_g}$ and $\omega_{+}\rightarrow\omega_q=\sqrt{k^2+\Pi_q}$ as in section 
\ref{QPM} with asymptotic masses $\Pi_g = \hat{m}_D^2/2$ and $\Pi_q = 2\hat{M}^2$, 
considering only one chemical potential $\mu_i=\mu_q$ ($\sum_i\mu_i^2 \rightarrow N_f\mu_q^2$). 
Integrating the logarithmic terms in (\ref{equ:phi27},~\ref{equ:phi28}) by 
parts (note that (\ref{equ:phi28+}) already obeys the desired form), one exactly recovers expressions 
(\ref{equ:entrcomp},~\ref{equ:denscomp}) of our QPM, where the replacement of $g^2$ by $G^2$ remains as phenomenological 
procedure on top of the listed "approximations". 

\section{Conclusions}
\label{cons}

In summary, motivated by the successful reproduction of available results of QCD thermodynamics, we attempted a 
collection of necessary steps to establish the link of our employed quasi-particle model to QCD. Quite severe assumptions 
had to be made. Even with these, resulting in the formal structure of our model, an additional and crucial point is 
the parametrization of the effective coupling. While allowing for an accurate two-parameter fit of many different 
lattice QCD results, it requires a foundation. In this respect, we refer to the work in \cite{Shuryak}, where 
the authors argue that the pure quasi-particle excitations, deduced from a preliminary study of the poles of quark and 
gluon propagators \cite{Petreczky}, are too heavy to saturate the pressure delivered from lattice calculations, 
e.~g. \cite{Kar2}, i.~e. signalling the necessity of including additional degrees of freedom. Further systematic studies 
of the relevant degrees of freedom in the strongly coupled quark-gluon medium near $T_c$ are highly awaited to have 
some guidance. 

An additional issue is the chiral extrapolation. The quark masses of $0.4\,T$ employed in the here analyzed lattice simulations 
correspond to unphysically heavy pions with $m_\pi \sim 770$ MeV. In the 1-loop and HTL gluon self-energy considered here, 
such a finite quark mass has a tiny (negligible) impact. A more rigorous treatment of finite quark mass effects must be 
accomplished to arrive at a suitable chiral extrapolation. 

\bigskip
\noindent
{\it Acknowledgements.} M.B. would like to thank the organizers of {\it Hot Quarks 2006} for 
the invitation to a stimulating workshop. This work is supported by BMBF 06 DR 121, GSI and Helmholtz association VI. 
We thank J.P. Blaizot, F. Karsch, E. Laermann and A. Peshier for fruitful discussions. 
 
%
%
%
%

\begin{thebibliography}{100}

\bibitem{Allton1} C.R. Allton et al., Phys. Rev. D {\bf 66}, 074507 (2002) 
\bibitem{FK} Z. Fodor, S.D. Katz, Phys. Lett. B {\bf 534}, 87 (2002) 
\bibitem{Fodor} Z. Fodor, S.D. Katz, K.K. Szabo, Phys. Lett. B {\bf 568}, 73 (2003) 
\bibitem{deForcrand} P. de Forcrand, O. Philipsen, Nucl. Phys. B {\bf 642}, 290 (2002); 
  Nucl. Phys. B {\bf 673}, 170 (2003) 
\bibitem{Lombardo} M. D'Elia, M.-P. Lombardo, Phys. Rev. D {\bf 67}, 014505 (2003); Phys. Rev. D {\bf 70}, 074509 (2004) 
\bibitem{Arnold95} P.~Arnold, C.~Zhai, Phys. Rev. D {\bf 51}, 1906 (1995) 
\bibitem{Zhai95} C.~Zhai, B.~Kastening, Phys. Rev. D {\bf 52}, 7232 (1995) 
\bibitem{Kajantie03} K.~Kajantie et al., Phys. Rev. D {\bf 67}, 105008 (2003) 
\bibitem{Vuorinen} A. Vuorinen, Phys. Rev. D {\bf 67}, 074032 (2003); Phys. Rev. D {\bf 68}, 054017 (2003) 
\bibitem{Ipp04} A. Ipp, A. Rebhan, A. Vuorinen, Phys. Rev. D {\bf 69}, 077901 (2004)
\bibitem{Levai} P.~Levai, U.~Heinz, Phys.~Rev. C {\bf 57}, 1879 (1998)
\bibitem{Schneider} R.A.~Schneider, W.~Weise, Phys. Rev. C {\bf 64}, 055210 (2001) 
\bibitem{Letessier} J.~Letessier, J.~Rafelski, Phys. Rev. C {\bf 67}, 031902 (2003)
\bibitem{Rebhan} A. Rebhan, P. Romatschke, Phys. Rev. D {\bf 68}, 025022 (2003)
\bibitem{Thaler} M.A.~Thaler, R.A.~Schneider, W.~Weise, Phys. Rev. C {\bf 69}, 035210 (2004)
\bibitem{Ivanov} Yu.B.~Ivanov, V.V.~Skokov, V.D.~Toneev, Phys. Rev. D {\bf 71}, 014005 (2005)
\bibitem{Peshier} A. Peshier et al., Phys. Lett. B{\bf 337}, 235 (1994); Phys. Rev. D {\bf 54}, 2399 (1996); 
  A. Peshier, B. K\"ampfer, G. Soff, Phys. Rev. C {\bf 61}, 045203 (2000); 
  Phys. Rev. D {\bf 66}, 094003 (2002)
\bibitem{Rischke} D.H. Rischke, Prog. Part. Nucl. Phys. {\bf 52}, 197 (2004) 
\bibitem{Andersen} J.O.~Andersen, E.~Braaten, M.~Strickland, Phys. Rev. Lett. {\bf 83}, 2139 (1999); 
Phys. Rev. D {\bf 61}, 014017 (1999); Phys. Rev. D {\bf 61}, 074016 (2000) 
\bibitem{Andersen02b} J.O.~Andersen, M.~Strickland, Phys. Rev. D {\bf 66}, 105001 (2002) 
\bibitem{Blaizot} J.P.~Blaizot, E.~Iancu, A.~Rebhan, Phys. Rev. Lett. {\bf 83}, 2906 (1999); 
Phys. Lett. B {\bf 470}, 181 (1999); Phys. Lett. B {\bf 523}, 143 (2001); in 
{\it Quark Gluon Plasma}, edited by R.C. Hwa (World Scientific, Singapore, 2003)
\bibitem{Blaizot01} J.P.~Blaizot, E.~Iancu, A.~Rebhan, Phys. Rev. D {\bf 63}, 065003 (2001)
\bibitem{Blaizot02} J.P.~Blaizot, E.~Iancu, Phys. Rept. {\bf 359}, 355 (2002)
\bibitem{Gorenstein} M.I. Gorenstein, S.N. Yang, Phys. Rev. D {\bf 52}, 5206 (1995)
\bibitem{LeBellac96} M.~Le~Bellac, {\it{Thermal field theory}} (Cambridge University Press,
  Cambridge, England, 1996)
\bibitem{Pisarski} R.D. Pisarski, Physica A {\bf 158}, 146 (1989); Nucl. Phys. {\bf A498}, 423c (1989)
\bibitem{Blu05} M.~Bluhm, B.~K\"ampfer, G.~Soff, Phys.~Lett. B {\bf 620}, 131 (2005)
\bibitem{Kar1} F. Karsch, E. Laermann, A. Peikert, Phys. Lett. B {\bf 478}, 447 (2000)
\bibitem{Kar2} F. Karsch, K. Redlich, A. Tawfik, Eur. Phys. J. C {\bf 29} (2003) 549 
\bibitem{Kar3} F. Karsch, Nucl. Phys. Proc. Suppl. {\bf 83}, 14 (2000) 
\bibitem{All05} C.R. Allton et al., Phys. Rev. D {\bf 68}, 014507 (2003); Phys. Rev. D {\bf 71}, 054508 (2005)
\bibitem{Luttinger60} J.M.~Luttinger, J.C.~Ward, Phys. Rev. {\bf 118}, 1417 (1960)
\bibitem{Baym62} G.~Baym, Phys. Rev. {\bf 127}, 1391 (1962)
\bibitem{Lee60} T.D.~Lee, C.N.~Yang, Phys. Rev. {\bf 117}, 22 (1960)
\bibitem{Kapusta89} J.I.~Kapusta, {\it{Finite-temperature field theory}} (Cambridge University Press,
  Cambridge, England, 1989)
\bibitem{Peshier01} A.~Peshier, Phys. Rev. D {\bf 63}, 105004 (2001); Nucl. Phys. A {\bf 702}, 128 (2002) 
\bibitem{Freedman78} B.A. Freedman, L. McLerran, Phys. Rev. D {\bf 16}, 1130 (1978)
\bibitem{Peshier98a} A.~Peshier et al., Eur. Phys. Lett. {\bf 43}, 381 (1998)
\bibitem{Vanderheyden} B.~Vanderheyden, G.~Baym, J. Stat. Phys. {\bf 93}, 843 (1998); in 
  {\it Progress in Nonequilibrium Green's functions}, edited by M. Bonitz (World Scientific, Singapore 2000)
\bibitem{Peshier04} A.~Peshier, Phys. Rev. D {\bf 70}, 034016 (2004); J. Phys. G {\bf 31} 371 (2005) 
\bibitem{Thoma} A. Peshier, K. Schertler, M.H. Thoma, Annals Phys. {\bf 266}, 162 (1998)
\bibitem{Kalashnikov} O.K. Kalashnikov, Fortsch. Phys. {\bf 32}, 525 (1984) 
\bibitem{Shuryak} E.V. Shuryak, I. Zahed, Phys. Rev. D {\bf 70}, 054507 (2004) 
\bibitem{Petreczky} P. Petreczky et al., Nucl. Phys. Proc. Suppl. {\bf 106}, 513 (2002)
%
%
\end{thebibliography}
%

\end{document}